\documentclass{article}[12]
\usepackage{amsmath}
\usepackage{amsfonts}
\usepackage{amssymb}
\usepackage{amsthm}
\usepackage{ifpdf}
\usepackage{subfig}
\usepackage{wrapfig}
\usepackage{fullpage}
\usepackage{cite}

\ifpdf

  \usepackage[pdftex]{epsfig}
  \usepackage[pdftex]{hyperref}

\else

    \usepackage[dvips]{epsfig}
    \newcommand{\href}[2]{#2}

\fi

\theoremstyle{definition}
\newtheorem{theorem}{Theorem}[section]

\begin{document}

\title{%Negative Interactions in Self-Assembly:  Oscillators, Temperature Programming, and Fuel Efficient Computation\\ OR \\
        %Dynamically Reprogrammable Self-Assembly Based Computer\\ OR \\
        Fuel Efficient Computation in Passive Self-Assembly}
\author{
Robert Schweller\thanks{Department of Computer Science, University of Texas - Pan American,
      \protect\url{rtschweller@utpa.edu} This author's research was supported in part by National Science Foundation Grant CCF-1117672.}
\and
Michael Sherman\thanks{Department of Computer Science, University of Texas - Pan American,
      \protect\url{mjsherman@utpa.edu} This author's research was supported in part by National Science Foundation Grant CCF-1117672.}
}
\date{}
\maketitle

\begin{abstract}
In this paper we show that passive self-assembly in the context of the tile self-assembly model is capable of performing fuel efficient, universal computation.  The tile self-assembly model is a premiere model of self-assembly in which particles are modeled by four-sided squares with glue types assigned to each tile edge.  The assembly process is driven by positive and negative force interactions between glue types, allowing for tile assemblies floating in the plane to combine and break apart over time.   We refer to this type of assembly model as passive in that the constituent parts remain unchanged throughout the assembly process regardless of their interactions.  A computationally universal system is said to be fuel efficient if the number of tiles used up per computation step is bounded by a constant.  Work within this model has shown how fuel guzzling tile systems can perform universal computation with only positive strength glue interactions~\cite{Winf98}.  Recent work has introduced space-efficient, fuel-guzzling universal computation with the addition of negative glue interactions and the use of a powerful non-diagonal class of glue interactions~\cite{DotKarMas10}. Other recent work has shown how to achieve fuel efficient computation~\cite{padilla2012ASP} within active tile self-assembly.  In this paper we utilize negative interactions in the tile self-assembly model to achieve the first computationally universal passive tile self-assembly system that is both space and fuel-efficient.  In addition, we achieve this result using a limited diagonal class of glue interactions.
\end{abstract}

\thispagestyle{empty}\newpage\setcounter{page}{1} \clearpage

\section{Introduction}
%Citations:
%
%Active Self-Assembly:
%
%Eric Klavins work? (maybe just Damien's paper as the most recent example of active self-assembly is enough).
%
%Damien's nubots papers, \cite{woods2012EAS}.
%
%cite the crystalline robots papers?
%
%
%self-assembly def, active and passive def

Self-assembly is the process by which systems of simple objects organize themselves through local interactions into larger, more complex objects.  There are perhaps two categories of self-assembly: passive and active.  In passive self-assembly the objects of the system are simple, stagnant particles that interact simply by surface chemistry and geometry.  In contrast, objects within an active self-assembly model may be permitted to move, rotate, adjust state, or add and remove bonding domains based on local interactions.  Put another way, active self-assembly extends traditional passive self-assembly by considering the objects of the system to be simple robots with abilities that vary according to the particular model considered.

The study of passive self-assembly is important for multiple reasons.  First, many theoretical models of active self-assembly do not currently have satisfactory implementations at the nanoscale.  In contrast, many computationally interesting passive tile self-assembly constructions are seeing experimental success based on DNA implementations~\cite{RoPaWi04, fuji2007TRA, BarSchRotWin09}.  Further, it is plausible that many active self-assembly models will see implementation through constructions built up within passive self-assembly models.  For example, the system proposed in this paper which implements a space-efficient, fuel efficient universal computation system might be modified to serve as the internal machinery for a larger, active component, thus potentially allowing for a passive self-assembly system to simulate the behavior of a more powerful active system.  Finally, it is important to understand the power and limits of passive self-assembly to better understand when active components are truly necessary.

In this paper we focus on an established model of passive self-assembly, the tile assembly model (TAM).  Monomers in the TAM are unit squares with glue types assigned to edges.  Self-assembly is driven by a large (infinite) number of copies of a set of tile types floating about, bumping into one another in the plane, and potentially sticking together when glue affinities exceed some set threshold.  While simple, the TAM has been extensively studied~\cite{AdChGoHu01, AdlCheGoeHuaWas01, ACGHKMR02, AKKRS09, AGKS05, BeckerRR06, KaoSchS08, KS06, DDFIRSS07, Sum09, Dot10, ChaGopRei09,RNaseSODA2010,SolWin07,DemPatSchSum2010RNase, ManStaSto09ISAAC, BryChiDotKarSek11PNSA, DotPatReiSchSum10, CooFuSch11, CheDot12, CheDotSek11,fu2012SAGT,demaine2011ODS,cannon2012two} and shown to be capable of universal computation~\cite{Winf98}, and even recently shown to be intrinsically universal~\cite{USAreal} in the case of the \emph{abstract} TAM.

Within the TAM, we consider the problem of improving existing universal computation results by obtaining \emph{fuel-efficiency}.  A TAM system that simulates a Turing machine is said to be fuel-efficient if only a constant number of tile types are used up per computation step.  This requirement, or close approximations to it, are of fundamental importance for the implementation of scalable molecular computers.  Unfortunately, all passive variants of the TAM have failed to yield fuel-efficient universal computation.

To address this fuel-\emph{deficiency}, we consider the TAM with the added power of negative glue interactions.  The power of negative force interactions stems from the possibility of stable assemblies coming together with strong affinities and yet producing unstable products that subsequently break apart into pieces that are different from the original pair of assemblies.  By careful engineering, this property can be harnessed to allow assemblies to attach to specific locations of large assemblies, surgically pry off and replace small portions of the assembly, and finally detach the replacement mechanism and repeat the process.  If you imagine the large assembly representing the tape of a Turing machine and the location of attachment is at the current head location, you get a high level overview of our Turing machine simulation that only uses up a fixed number of tile types per computation step.  This stands in contrast to previous fuel guzzling work in which each computation step is simulated by the assembly of a complete, slightly modified, copy of the entire tape.

\subsection{Related Work}
%\begin{itemize}
%    \item Main one: Negative interactions in irreversible self-assembly \cite{DotKarMas10}
%    \item Fuel efficient turing machines:  signal tiles \cite{padilla2012ASP}
%    \item bit flipping: temperature programming papers, mine and scott's \cite{KS07,Sum09}
%    \item breaking apart self-assembly: RNA papers like shape replication \cite{RNaseSODA2010,RNAPods}
%    \item maybe damiens nubots if its available as example of detachable assemblies \cite{woods2012EAS}
%\end{itemize}

Doty~et.~al.~\cite{DotKarMas10} first considered the effect of negative interactions in self-assembly and showed how to design space efficient computational self-assembly systems by utilizing negative glues to slice off a previous assembly state after computing the next state.  This state transition construction can be applied to the simulation of a Turing machine.  However, their construction does not achieve fuel efficiency in that each computational step uses up a number of tiles on the order of the current size of the tape.  Their result is also not applicable to our graph traversal system in that their technique, modified to the graph traversal problem, would lead to dead end assemblies that violate the required dynamics of the graph traversal.  Finally, their paper utilizes a very powerful \emph{non-diagonal} version of the tile self-assembly glue function.  We restrict ourselves to the much more restricted diagonal class of glue functions which is likely more feasible for experimental implementation.  Additionally, Doty~et.~al. provide an elegant amortized analysis proof showing a limit to the extent to which tile reuse can be achieved with negative force interactions within the TAM.  We implicitly use their result in that it allows us to reasonably simplify our definition of fuel usage to not deal with tile reuse.

Some recent work has been done that shows provable increases in power of active self-assembly systems over passive TAM systems.  Recent work by Woods~et.~al.~\cite{woods2012EAS} has shown that local rule based active self-assembly systems are able to assemble large shapes in poly-logarithmic time when the monomers of the system are able to perform operations such as pushing a row of placed monomers a unit of distance in some direction.  Another recent work by Padilla~et.~al.~\cite{padilla2012ASP} considers an active tile self-assembly model motivated by a DNA strand replacement mechanism in which tiles can pass simple, fire-once signals from one tile edge to another.  They show that this simple mechanism allows for the implementation of fuel-efficient universal computation, as well as additional efficiencies that are provably impossible within the passive TAM.

Additional investigations have looked into self-assembly models that allow self-assembly detachment, a key requirement for the implementation of space efficient computations~\cite{AGKS05,KS06,Sum09,RNaseSODA2010}.  Other recent work has considered negative glues within the TAM to achieve universal computation with a very limited set of glues at temperature-1~\cite{rgTAM}.

\paragraph{Paper layout.}  In Section~\ref{sec:definitions} we define the tile assembly model, as well as present definitions that define our graph traversal and Turing machine simulation problems.  In Section~\ref{sec:graphWalking} we present our construction for fuel efficient graph walking.  Finally, in Section~\ref{sec:turingmachine}, we extend our graph walking construction to simulate Turing machines.

\section{Definitions and Model}\label{sec:definitions}
\begin{figure}[htp]
	\begin{center}
	\includegraphics[scale=1.0]{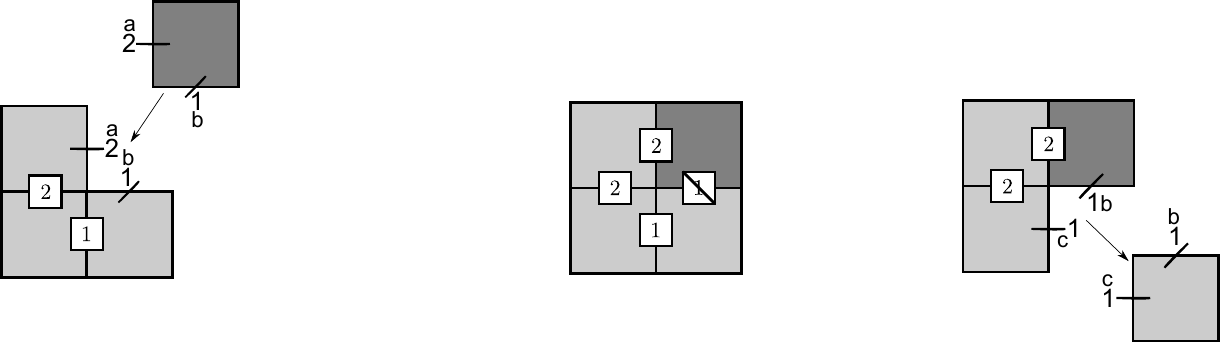}
	\caption{ This figure introduces our notation for our constructions, as well as a simple example of combination and detachment events.  We denote positive glues with vertical and horizontal lines protruding from tile edges along with numbers and labels denoting glue strength and type.  Negative strength glues are labeled with slanted lines.  Finally, bonded glues between adjacent tiles are depicted with a number denoting the bond strength inscribed within a box between the bonded tile edges.  (a) The 3-tile assembly on the left and the singleton gray tile are combinable into a $2\times 2$ square shown in (b) as the cut between the 2 assemblies yields strength 1.  Note, however, that the producible $2\times 2$ square in (b) is not stable and is breakable into the assemblies shown in (c).
	\label{fig:modelex} }
	\end{center}
\end{figure}

In this section we first define the two-handed tile self-assembly model with both negative and positive strength glue types and diagonal glue functions.  We also formulate the problem of designing a tile assembly system that walks a given input graph, as well as the concept of fuel efficiency.

\subsection{Tile Self-Assembly Model}
\paragraph{Tiles.}  Consider some alphabet of glue types $\Pi$.  A tile is a finite edge polygon with some finite subset of border points each assigned some glue type from $\Pi$.  Further, each glue type $g \in \Pi$ has some rational number strength $str(g)$.  Finally, each tile may be assigned a finite length string \emph{label}.  In this paper we consider a special class of tiles that are unit squares of the same orientation with at most one glue type per face, with each glue being placed exactly in the center of the tile's face.

\paragraph{Assemblies.}  An assembly is a set of tiles whose interiors do not overlap.  For a given assembly $\Upsilon$, define the \emph{bond graph} $G_\Upsilon$ to be the weighted graph in which each element of $\Upsilon$ is a vertex, and each edge weight between tiles is the sum of the strengths of the overlapping, matching glue points of the two tiles.  An assembly $C$ is said to be \emph{stable} if the bond graph $G_C$ has min-cut at least 1, and \emph{unstable} if the min-cut is less than 1. Note that if the set of border points of all tiles in an assembly is not a connected set, then the assembly cannot be stable.  Note that only overlapping glues that are the same type contribute a non-zero weight, whereas overlapping, non-equal glues always contribute zero weight to the bond graph.  The property that only equal glue types interact with each other is referred to as the \emph{diagonal glue function} property and is perhaps more feasible for experimental implementation than more general glue functions.  Additionally, with the square tiles considered in this paper, stable assemblies will necessarily consist of tiles stacked face to face, forming a subset of the 2D grid.

For an assembly $A$, let $A^*$ denote the set of all assemblies that are equal to $A$ up to translation.  For a set of assemblies $T$, let $T^*$ denote the set of all assembly sets $A^*$ such that $A\in T$.  Put another way, $T^*$ is a set that represents a set of assemblies $T$ if we do not care about translation.

%For an assembly $C$ whose bond graph $G_C$ has min-cut at least 1, we say that $C$ is \emph{stable}.  An assembly whose bond graph has a min-cut less than 1.  Note that any assembly in which the set of border points of all tiles in the assembly is not a connected set will not be stable.

\paragraph{Breakable Assemblies.}
For an (unstable) assembly $C$ whose bond graph $G_C$ has a cut into assemblies $A$ and $B$ with weight less than 1, we say that $C$ is \emph{breakable} into $A$ and $B$.

\paragraph{Combinable Assemblies.}
Informally, two assemblies $A$ and $B$ are said to be combinable into an assembly $C$ if the assemblies can be translated together in such a way that they do not overlap and the sum of the matched glue strengths between the two assemblies is at least 1. Formally, consider two assemblies $A$ and $B$.  If $B$ can be translated into $B'$ such that $C= A \bigcup B'$ is a valid (not overlapping) assembly such that the cut of $G_{A\bigcup B'}$ into $A$ and $B'$ has strength at least 1, then we say that $A$ and $B$ are \emph{combinable} into $C$.

Note that $A$ and $B$ may be combinable into an assembly that is not stable.  This is a key property that is leveraged throughout our constructions.  See Figure~\ref{fig:modelex} for an example.

\paragraph{Producible Assemblies.}
A set of initial assemblies $T$ has an associated set of \emph{producible} assemblies, $\texttt{PROD}_T$, which define what assemblies can grow from the initial set $T$ by any sequence of combination and break events. Formally, $T \subseteq \texttt{PROD}_T$ as a base case set of producible assemblies.  Further, given any $A,B \in \texttt{PROD}_T$ with $A$ and $B$ combinable into $C$, then $C \in \texttt{PROD}_T$, and for any $C \in \texttt{PROD}_T$ where $C$ is breakable into $A$ and $B$, then $A,B \in \texttt{PROD}_T$.  Put another way, the set of producible assemblies contains any assembly that can be obtained by a valid sequence of combinations and breaks, starting from the initial set of assemblies $T$.  We further define the set of \emph{terminal} assemblies, $\texttt{TERM}_T$, to be the subset of $\texttt{PROD}_T$ such that no element of $\texttt{TERM}_T$ can undergo a break or combination transition with another element of $\texttt{PROD}_T$.  Note that $\texttt{PROD}_T = \bigcup_{X\in \texttt{PROD}^*_T} X$ and $\texttt{TERM}_T= \bigcup_{X\in \texttt{TERM}^*_T} X$.

Typically, we require that an initial assembly set consist only of stable assemblies.  When $T$ consists only of assemblies that are singleton tiles, $T$ is called a \emph{tile set}.

%\paragraph {What is a labeled tile and assembly.} Probably include this stuff under Tiles and assemblies... In the tile description include the ability for it to carry a label. Define the label of an assembly somehow for detachments and attachments... i.e. if a tile attaches with a label then the assembly inherits that label. The label of an assembly is any of the labels of its children

\subsection{Additional Definitions}

\paragraph{Valid Assembly Sequence.}  For an initial assembly set $T$, a valid assembly sequence for $T$ is any sequence of assemblies $S=\langle a_1, \ldots a_k \rangle$ such that each $a_i \in \texttt{PROD}_T$, and for each $i$ from 1 to $k-1$, $a_i$ is either combinable with some $b \in \texttt{PROD}_T$ to form $a_{i+1}$, or $a_i$ is breakable into $a_{i+1}$ and $b$ for some assembly $b$.  In addition, a valid assembly sequence $S$ is said to be \emph{$\ell$-focused} for label $\ell$ if each assembly in $S$ contains at least one tile with label $\ell$.  A valid assembly sequence is said to be \emph{nascent} if $a_1 \in T$.

For a given partial function $f: \texttt{PROD}_T \rightarrow V$ for a set $V$, define function $f'( \langle a_1,\ldots a_k \rangle )$ to be a function from assembly sequences over $\texttt{PROD}_T$ to sequences over $V$ defined by replacing each $a_i$ with $f(a_i)$ if $f(a_i)$ is defined, and deleting $a_i$ if $f(a_i)$ is not defined.

\paragraph{Graph Walking Assemblies.}  Consider a directed graph $G=(V,E)$, a start vertex $s\in V$, and an initial assembly set $T$ in which some tiles of some assemblies are labeled with $\ell$.  $T$ is said to \emph{walk} graph $G$ if there exists a partial function $f: \texttt{PROD}_T \rightarrow V$ such that:

\begin{enumerate}
    \item (assembly sequences are walks) For any $\ell$-focused nascent assembly sequence $S$ of $T$, $f'(S)$ is a valid walk of $G$ starting at vertex $s$.
    \item (walks are assembly sequences) For any walk $W$ of $G$ starting at vertex $s$, there exists a nascent, $\ell$-focused assembly sequence $S$ such that $f'(S) = W$.
    \item (no undesired dead-ends) Consider an edge $(u,v)\in E$.  Then for any $\ell$-focused assembly sequence $S$ such that $f'(S) = \langle u \rangle$, there exists an $\ell$-focused assembly sequence $R$ such that $S$ is a prefix of $R$ and $f'(R) = \langle u, v \rangle$.
\end{enumerate}

In this definition, the $\ell$ label is meant to denote tiles and assemblies that are meant to be part of the output, and not garbage.  That is, to walk a graph efficiently there will necessarily be dead-end garbage assemblies.  Our definition thus requires that all such garbage assemblies be free of the $\ell$ label and thus can be distinguished from the assemblies that represent the graph walking assemblies. In a sense, our definition states that if you stare at one $\ell$-labeled tile within an assembly, over time the assembly containing that tile will be guaranteed to walk the graph correctly, whereas if you stare at a portion of the assembly that is not labeled with the $\ell$, it is possible that portion may detach and become garbage at some point.

\paragraph{Fuel Efficiency.} \label{sec:fuelefficient}
Consider two assemblies $a$ and $b$.  Define $fuel(a,b)$ to be $|b| - |a|$ if $|b| \geq |a|$, and zero otherwise.  For an assembly sequence $A = \langle a_1, \ldots a_k \rangle$, define $fuel(A)$ to be $\sum fuel(a_i, a_{i+1})$.

For a tile set $T$ that walks a directed graph $G$, $T$ is said to be \emph{fuel efficient} if for all $\ell$-focused, nascent assembly sequences $S$, $fuel(S)/|f'(S)| = O(1)$.  That is, for all possible walks, the average fuel per vertex transition is bounded by a constant.

\section{Graph Walking}\label{sec:graphWalking}

In this section we construct a tile assembly set for the fuel-efficient walk of a given input directed graph:

\begin{theorem}\label{thm:graph}
For any given finite directed graph $G=(V,E)$, there exists a fuel efficient initial assembly set $T$ that walks $G$.  Further, $T$ contains at most $O(|V| + |E|)$ distinct assemblies each of $O(1)$ size.
\end{theorem}

This theorem can be slightly strengthened in that it is possible to modify the construction so that $T$ consists of only singleton tiles.  However, for clarity we do not present this modification.

The construction for this result serves two purposes.  First, we believe the result itself is interesting.  One example corollary is the ability to implement oscillator self-assembly systems with an example given in Section~\ref{sec:oscillator}.  The second purpose is that this construction presents some of the key techniques in a simplified form that we will extend to construct our main result: fuel efficient Turing machines.

The remainder of this section describes the construction for the proof of this theorem.

%\paragraph{Non-deterministic Walk.} A valid non-deterministic walk of graph $G$ starting at some vertex $v$ can be done with tile set $T$ and a partial mapping function $f : \texttt{PROD}_T \to V$ if the following constraints are met.
%
%\begin{description}\label{sec:constraints}
%\item[\quad $\bullet$ All Walks Begin With $v$:] $\forall s \in T_{RAS}$ the $0^{th}$ element of $F(s)$ must be equal to $v$
%\item[\quad $\bullet$ Every $s \in T_{RAS}$ Represents A Valid Walk:] $\forall s \in T_{RAS} \exists w \in G$ s.t. $F(s) = w$
%\item[\quad $\bullet$ All Walks Of $G$ Represented In $T_{RAS}$:] $\forall w \in G \exists s \in T_{RAS}$ s.t. $F(s) = w$
%\item[\quad $\bullet$ No Dead Ends:] It is required that $\forall s \in T_{RAS}$ if $F(s)$ ends on a vertex with outgoing edges, then $s$ must be a prefix of another sequence $r \in T_{RAS}$ such that $F(r)$ is at least one step longer than the walk of the graph $F(s)$ represents.
%\end{description}

%\subsection{Solution Overview}
%\begin{theorem}\label{thm:graph}
%There exists a fuel efficient, space efficient, step efficient tile set such that given a directed graph $G$ and a starting vertex $v$ the tile set non-deterministically traverses $G$.
%\end{theorem}
%
%The proof for Theorem \ref{thm:graph} follows by construction in sections \ref{sec:tileset} and \ref{sec:ocorrect}.

\subsection{Template Assembly Set}\label{sec:tileset}
The following tile assembly template allows one to create an assembly set that can non-deterministically walk across a directed graph. This template set creates an assembly set that has $O(|V| + |E|)$ assembly types which also uses only a small constant number of fuel tiles for each transition, nine. An example of constructing an assembly set using this template follows in Section \ref{sec:oscillator} where we build a base four oscillator.

\begin{figure}[htp]
	\subfloat[This tile allows an edge gadget to detach the it's corresponding vertex.] {
		\centering
		\parbox{1.5in} {
			\centering
			\includegraphics[scale=1.2]{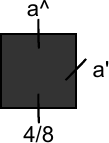}
			\label{fig:tiledetacher}
		}
	}
	~~~~~~
	\centering
	\subfloat[These tiles together represent a vertex in the graph.] {
		\centering
		\parbox{1.5in} {
			\centering
			\includegraphics[scale=1.0]{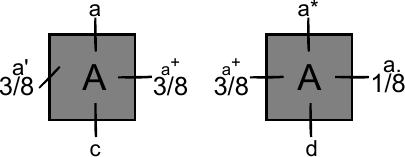}
			\label{fig:vertex}
		}
	}
	~~~~~~
	\subfloat[This tile allows an edge gadget to know when it's vertex has finished attaching.] {
		\centering
		\parbox{1.5in} {
			\centering
			\includegraphics[scale=1.2]{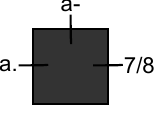}
			\label{fig:tileattached}
		}
	}
	\caption{These four tiles are needed in order to represent any vertex in a graph. All glues labeled with some form of an $a$ require unique glues to the specific vertex these tiles represent and may not be shared amongst other vertices.}
	\label{fig:vertextiles}
\end{figure}

\paragraph{Vertex Representation.}  For each vertex in the graph you must have the set of two tiles shown in Figure \ref{fig:vertex}, a vertex detachment tile shown in Figure \ref{fig:tiledetacher}, and finally a vertex attachment tile which is shown in Figure \ref{fig:tileattached}. The five glues labeled $a'$, $a^*$, $a.$, $a^+$ and $a$ on the duple of Figure~\ref{fig:vertex} are unique to some vertex $a$ while the two southern glues are the same for all pairs of tiles.

\begin{figure}[htp]
	\centering
	\subfloat[This assembly, edge gadget, represent a directed edge from vertex {$a$} to vertex {$b$}.] {
		\centering
		\includegraphics[scale=1.0]{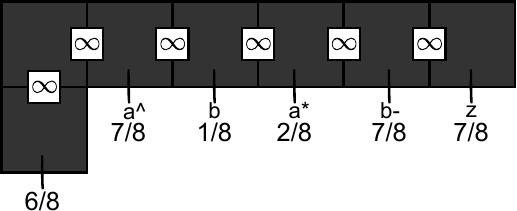}
		\label{fig:edge}
	}
	~~~~~~~
	\subfloat[This tile allows any edge gadget to detach itself once done.] {
		\centering
		\parbox[t]{1.5in} {
			\centering
			\includegraphics[scale=1.0]{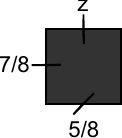}
			\label{fig:gadgetdetacher}
		}
	}
	\caption{The edge gadget and it's detachment tile.}
	\label{fig:detachmentstuff}
\end{figure}

\paragraph{Edge Gadget.}  Figure \ref{fig:edge} is the edge gadget which exchanges a vertex $a$ with the vertex $b$ on the cell assembly. When the edge gadget detaches itself from the cell assembly all tiles that detach with the gadget are no longer usable. In Figure \ref{fig:edge} the glues labeled with some $a$ represent glues unique to some vertex $a$ while the glues labeled with some $b$ represent glues specific to some vertex $b$. The glue labeled with a $z$ is unique to the edge detachment tile shown in Figure \ref{fig:gadgetdetacher}. This tile allows the edge gadget including it's junk to detach from the cell assembly.

\begin{figure}[htp]
	\center
	\includegraphics[scale=1.0]{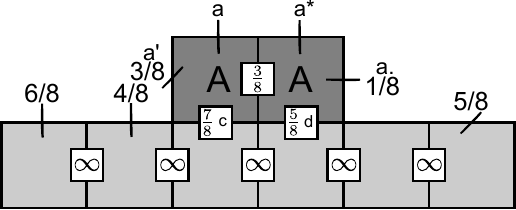}
	\caption{ The initial pre-built starting cell assembly with a vertex already attached.
	\label{fig:initial} }
\end{figure}
\paragraph{Cell Assembly.}  This cell assembly is assumed to be pre-built and contains the initial starting vertex $v$ already placed on top of the cell assembly. As you can see in Figure \ref{fig:initial} the glues $c$ and $d$ are not discriminatory and allow any two tiles representing a vertex to attach to the cell assembly with the aid of an edge gadget. The $\frac{6}{8}^{th}$ glue along with the $a*$ glue allow the edge gadget to attach. After the attachment the $\frac{4}{8}^{th}$ glue allows for the widget to get a better ``grip" on the assembly cell so that it may detach the current vertex. The final $\frac{5}{8}^{th}$ glue on the assembly allows the edge gadget to pry itself off.

\subsection{Correctness}\label{sec:ocorrect}

\begin{figure}[htp]
	\begin{center}
	\includegraphics[scale=0.3]{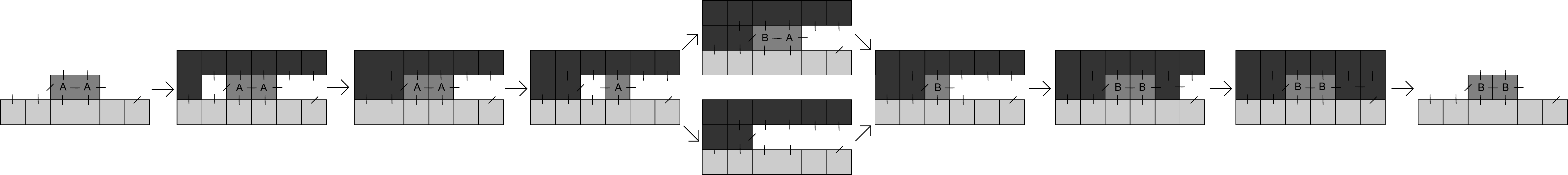}
	\caption{ The steps necessary to transition from one state to another.
	\label{fig:transition} }
	\end{center}
\end{figure}

As can be seen from Figure \ref{fig:transition} the possible transitions made by changing from one state to another are straight forward and verifying that each single tile addition must happen in the order shown is left as a short exercise to the reader. We will concentrate on the two negative glue detachment events which are more complex and need to be verified so as to ensure that no other cuts in the produced assemblies are unstable other than those depicted. We shall assume that each vertical column of the image represents a stage in the transition and so there are a total of $9$ stages from start to finish.

\subsubsection{Correct Gadget Removal}
After the gadget is done a tile may attach to the negative $\frac{5}{8}^{th}$ glue on the cell assembly which in turn pops off the edge gadget along with any junk. Two scenarios that are brought about by the infinite binding within the cell assembly and edge gadget independently. The first scenario occurs when the cell assembly and edge gadget end up on opposite sides of the detachment. The other happens when both the cell assembly and edge gadget end up on the same side of the detachment. We will analyze both these scenarios and show that from stage $8$ in Figure \ref{fig:transition} the only valid detachment is what results in stage $9$.

Suppose both the cell and the gadget detach to opposite sides of the cut. We must include a negative glue to get an unstable cut and so the edge detachment tile must be on the same side as the gadget. This gives us a cut with weight $\frac{1}{8}$ not taking into account any of the other four labeled tiles. For the cut to be unstable we can only increase the weight by $\frac{6}{8}$ so as to keep it below $1$. To stay unstable both the vertex detachment tile and the vertex attached tile must be on the side of the edge gadget because of their $\frac{7}{8}$ attachment. Also, the left vertex tile must stay on the side of the cell assembly for the same reason. With this new information the total weight of the cut is now $\frac{6}{8}$ and we only have the right vertex tile to place. The right vertex tile must now stay on the side of the cell because to keep the cut unstable we may only increase the weight of the cut by $\frac{1}{8}$ more. Therefore, it is easy to see that the only possible unstable cut if both the cell and gadget are on opposite sides is the one displayed in Figure \ref{fig:transition}.

Now, suppose that both the cell assembly and the edge gadget detach to the same side of the cut. In order to include the negative strength of the edge detachment tile it must be on the opposite side of both the edge and the gadget. This immediately shows us that our cut is $\frac{9}{8}$ which is stable. With no other negative glue interactions, there is no cut with both the cell assembly and the edge gadget on the same side that would lead to a detachment.

Therefore, since we have exhaustively explored both scenarios using the fact that the cell assembly and edge gadget are infinitely bound we can say that the only unstable cut at stage $8$ is the one that results in stage $9$.

\subsubsection{Correct Vertex Replacement}
Assume we are in stage $3$ of Figure \ref{fig:transition}.  We will show that the transition to stage $4$ is the only valid detachment event that may happen. If the cell and edge gadget were to detach to opposite sides of the cut this would give us a cut with weight $\frac{5}{8}$. Then, no matter which side of the cut you place the vertex removal tile all cuts will be above $1$. If both the edge gadget and the cell are on the same side of the cut, the left vertex tile must be on the opposite side of the cut since the vertex removal tile will not fall out. This means that either the left vertex tile or both vertex tiles must be on the opposite side of the cut. If both vertex tiles are on the opposite side of the cut we have a cut with weight $\frac{11}{8}$ which is stable. On the other hand, if it is only the left vertex tile on the opposite side of the cut then the weight of the cut is $\frac{7}{8}$ which enables it to detach. Therefore, the only detachment event that can occur after stage $3$ is the single left vertex tile detaching from the assembly leaving the assembly in stage $4$. Once in stage $4$ the right vertex tile destabilizes and may detach leaving room for the next vertex to attach and stage $6$ to come about.

\section{Quaternary Oscillator}\label{sec:oscillator}
\begin{figure}[htp]
	\begin{center}
	\includegraphics[scale=0.5]{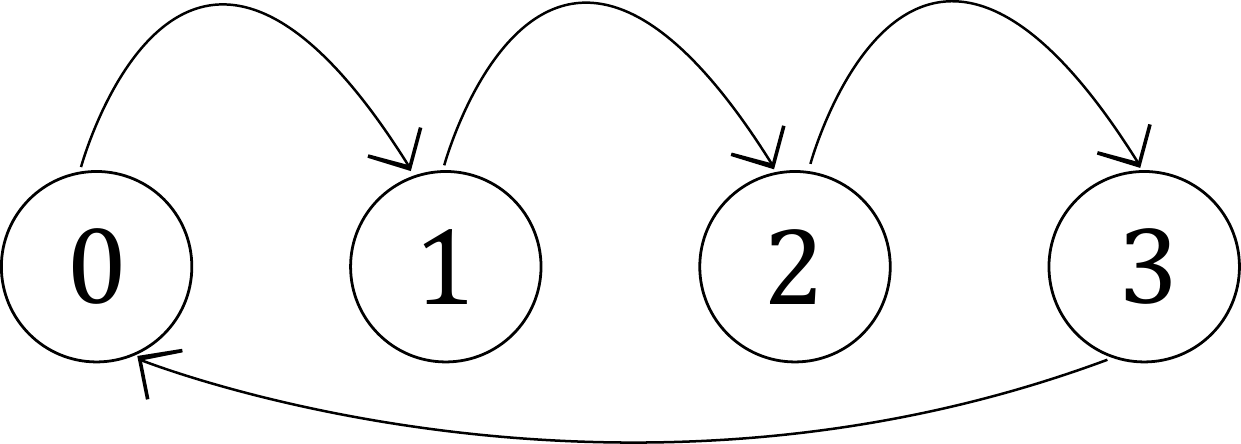}
	\caption{ The directed graph of a quaternary oscillator.
	\label{fig:base4graph} }
	\end{center}
\end{figure}

\begin{figure}[htp]
	\begin{center}
	\includegraphics[scale=.75]{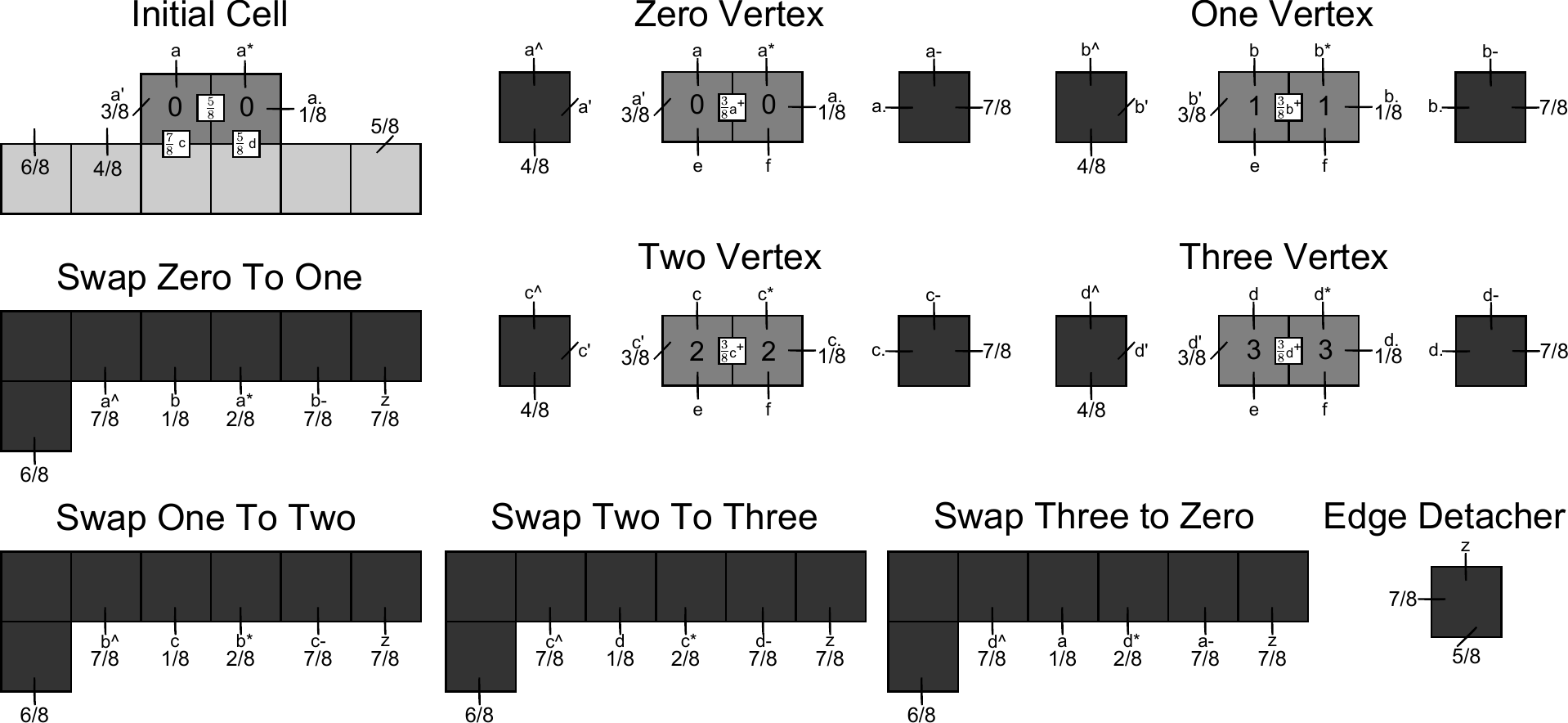}
	\caption{ The complete assembly set for creating a quaternary oscillator.
	\label{fig:base4tileset} }
	\end{center}
\end{figure}

A quaternary oscillator can be drawn as a directed graph with vertices representing each digit in the base-4 numeral system and the directed edges marking the transition from one numeral to the next. Figure \ref{fig:base4graph} shows a directed graph representing a quaternary oscillator. As can be seen, every vertex has exactly one outgoing edge and as such even if our Graph Walking Template Assembly Set is non-deterministic there is only one valid walk through the graph making the assembly set deterministic. Figure \ref{fig:base4tileset} represents the complete assembly set necessary to represent the graph in Figure \ref{fig:base4graph}. This is a straightforward example of how to create a concrete assembly set using the assembly template presented in Section \ref{sec:tileset}.

\section{Fuel Efficient Universal Computation} \label{sec:turingmachine}

%[Definition: We define the status $S$ of a Turing Machine $M$ as a 3-tuple $S = \langle state, tape, head \rangle$ where $state$ is the state of $M$, $tape$ represents the current tape, and $head$ represents the position of the head.]

In this section we describe how to construct an initial assembly set $T$ that will simulate a given Turing machine $M$.  To formally define simulation of a Turing machine, we utilize our graph walking formulation. For a given Turing machine $M$, we consider all possible tape configurations, states, and head locations the machine may enter over time and form a directed graph over these possible configurations depicting which configurations the machine may transition between within a single computation step.  With this definition we say that an assembly set $T$ simulates $M$ if it walks this configuration graph.

Formally, we define the simulation as follows:

\paragraph{Turing Machine Simulation.}  Consider some Turing Machine $M = \langle Q, \Gamma, b, \Sigma, \delta, q_0, F \rangle$, an initial tape $t$, and head position $h$.  Let a possible \emph{status} of $M$ be a 3-tuple $S = \langle \texttt{state}, \texttt{tape}, \texttt{head} \rangle$ where $\texttt{state}$ represents some current state from $Q$, $\texttt{tape}$ is some string of symbols over $\Gamma$, and $\texttt{head}$ specifies a location on the tape where the Turing machine head currently sits.  Additionally define the \emph{start} status for a Turing machine $M$ and an input tape $r$ as the status corresponding to $\texttt{tape} = r$, $\texttt{head} =h$, and $\texttt{state} =q_0$.

Given a Turing machine $M$, define a directed graph $G_M$ where the set of all statuses for $M$ are vertices, and there exists an edge from any status $a$ to status $b$ if and only if $M$ can transition from status $a$ to $b$ in one step.  Finally, we say an initial assembly set $T$ \emph{simulates} Turing machine $M$ if $T$ walks $G_M$.  Further, we say $T$ is a $\emph{fuel-efficient}$ simulation of $M$ if $T$ constitutes a fuel-efficient walk of $G_M$.

Our main result is as follows:

\begin{theorem}\label{thm:turing}  For any Turing machine $M$, there exists an initial assembly set $T$ such that $T$ is a fuel-efficient simulation of $M$.  Further, $|T| = O( |\Gamma| \cdot |Q| )$, and each element of $T$ is of $O(1)$ size.
\end{theorem}

As with our graph walking theorem, it is possible to strengthen this theorem such that $T$ contains only singleton tile assemblies.  For clarity of presentation and space, we omit the extended construction that achieves this.

The remainder of this section presents the construction that proves Theorem~\ref{thm:turing}

% then the initial status of the Turing Machine is $s = \langle q_0, t, h \rangle$. Now assume all possible status of $M$ are each represented by some vertex $v \in G_M$.  The edge $(a,b)$ is in $G_M$ if $M$ can go from status $a$ to status $b$ using a single computation step. We say that a tile system $T$ simulates a Turing Machine $M$ if $T$ walks $G_M$ and the initial vertex for the walk represents $s$.

\subsection{Template Tile Set}

\begin{figure}[htp]
	\begin{center}
	\includegraphics[scale=1.2]{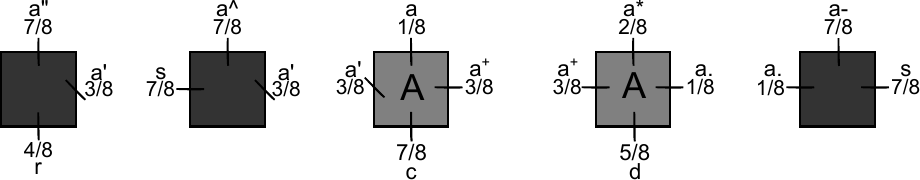}
	\caption{ These tiles represent a symbol and state pair along with their needed utility tiles.
	\label{fig:statesymbol} }
	\end{center}
\end{figure}
\paragraph{Symbols And States.} For our construction we add an additional state, the empty state $\varnothing$, to $M$ which will be used to represent a symbol with no state i.e. a location where the head of the Turing Machine currently is not. For every symbol in $\Gamma$ and every state in $Q$ including $\varnothing$ we create the set of tiles shown in Figure \ref{fig:statesymbol}. This means that we should have exactly $|\Gamma| \cdot (|Q| + 1)$ sets of tiles. The two center tiles labeled with an $A$ in Figure \ref{fig:statesymbol} will represent some duple $(a \in Q, s \in \Gamma \cup \varnothing)$. We define a stateless state symbol pair as having the $\varnothing$ state while a stateful state symbol pair has some $q \in Q$. The two leftmost tiles are called state symbol detachment tiles and detach the particular state symbol tile pair they were created with. This will be useful later when trying to transition the state of the Turing Machine. Finally, the last tile on the right is the state symbol attached tile and will allow the assemblies to know that the state symbol tile pair associated with it have both been attached.

%%%%%%%%%%%%%%%%%%%%%%%%%%%%%%%%%%%%%%%%%%%%%%%%%%%%%%%%%%%%%%%%%%%%%%%%%%%%%%%%%%%%%%%%%%%%%%%%%%%%%%%%%%%%%%%

\begin{figure}[htp]
	\begin{center}
	\includegraphics[scale=1.2]{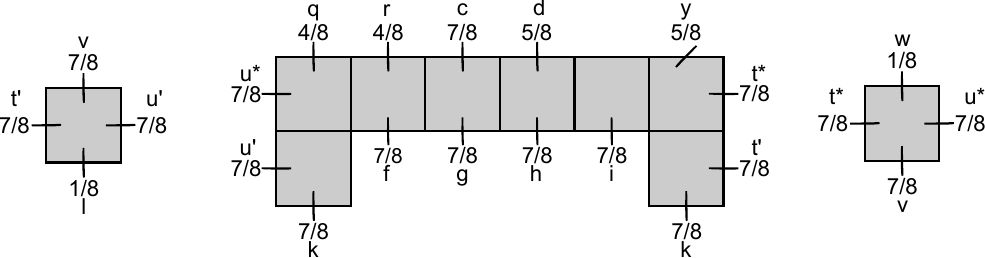}
	\caption{ This assembly represents a single position on the tape. It is a tape cell which can and should contain some symbol state pair atop it attached to glues $c$ and $d$. The two accompanying tiles are what will allow the tape to extend to the right or left.
	\label{fig:tapecell} }
	\end{center}
\end{figure}
\paragraph{Tape Cell.}  A tape cell in our construction is a single piece of tape which contains some state symbol tile pair taken from Figure \ref{fig:statesymbol}. If the state contained on the tape cell is $\varnothing$ then the tape cell contains only data, i.e. a symbol from $\Gamma$. On the other hand, if the state contained on the tape is not $\varnothing$ then the location of the Turing Machine's head is at that location and it contains some symbol from $\Gamma$. The whole of the tape is then at least one tape cell possibly combined with many more tape cells attached together with the head of the tape in only a single location i.e. with a single cell containing a state unequal to $\varnothing$ at any given time. Should the head of the Turing Machine be placed in multiple locations it will result in undefined behavior. When we later extend the tape we insure that the extended piece of tape gets initialized automatically to the blank state; i.e. the symbol $b$ in combination with the state $\varnothing$.

%%%%%%%%%%%%%%%%%%%%%%%%%%%%%%%%%%%%%%%%%%%%%%%%%%%%%%%%%%%%%%%%%%%%%%%%%%%%%%%%%%%%%%%%%%%%%%%%%%%%%%%%%%%%%%%

\begin{figure}[htp]
	\begin{center}
	\includegraphics[scale=1.1]{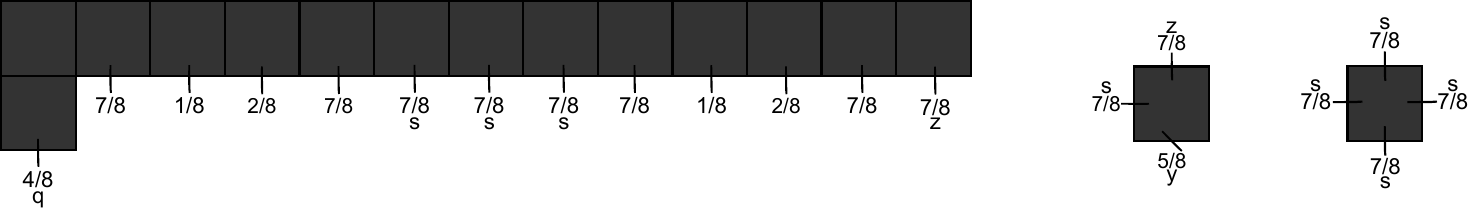}
	\caption{ The leftmost assembly represents a single ``Turing Instruction" and is called a transition gadget. The second tile is the detachment tile and the last tile is simply a tile that will be used to fill space.
	\label{fig:turinginstruction} }
	\end{center}
\end{figure}

\paragraph{Transition Function.}  A Turing Instruction is the 5-tuple $I = \langle c \in Q, s \in \Gamma, p \in \Gamma, \{L, R\}, n \in Q \rangle$ where $c$ is the current state, $s$ is the scanned symbol, $p$ is the symbol to print, $\{L, R\}$ is the left or right movement of the head, and $n$ is the new state. For each Turing Instruction in $M$ create $|\Gamma|$ transition gadgets. A single transition gadget is shown in Figure \ref{fig:turinginstruction}.

%We have already said that we will use the two center tiles in Figure \ref{fig:statesymbol} to represent a state symbol pair. This pair has two detachment tiles and a completion tile which are also shown in Figure \ref{fig:statesymbol} as the left most two tiles and rightmost tile respectively. The difference between the edge gadget and the transition gadget is that instead of exchanging a single pair of tiles we exchange two pairs of tiles. The extension from a single vertex to two state symbol pairs is rather straight forward. There are five steps necessary to change a vertex in the Graph Walking Template Tile Set. The first step is to attach to the correct vertex. Then we need to detach the current vertex and attach the next vertex. Following that we must be sure that the next vertex has properly attached and finally detach the gadget responsible for this transition.

For two state symbol pairs we follow the same process as exchanging a single vertex in the Graph Walking Template with a few exceptions. The most obvious exception is that we are now exchanging two symbol state pairs instead of a single vertex pair. In order to facilitate this change we extend the transition gadget to accompany the two exchanges. The next difference is that after exchanging the first pair of tiles the next symbol state pair must be exchanged instead of detaching the whole transition gadget.  This is the reason for needing two state symbol detachment tiles instead of just the single one as in the Graph Walking Template.  Now, of the two detachment tiles the first is used when detaching the left state symbol pair underneath the transition gadget and the second is used when detaching the right state symbol pair underneath the transition gadget. This along with the changing of a couple glue strengths gives us a transition gadget shown in Figure \ref{fig:turinginstruction} that is able to exchange two symbol state pairs at the same time.

\begin{figure}[htp]
	\begin{center}
	\includegraphics[scale=1.0]{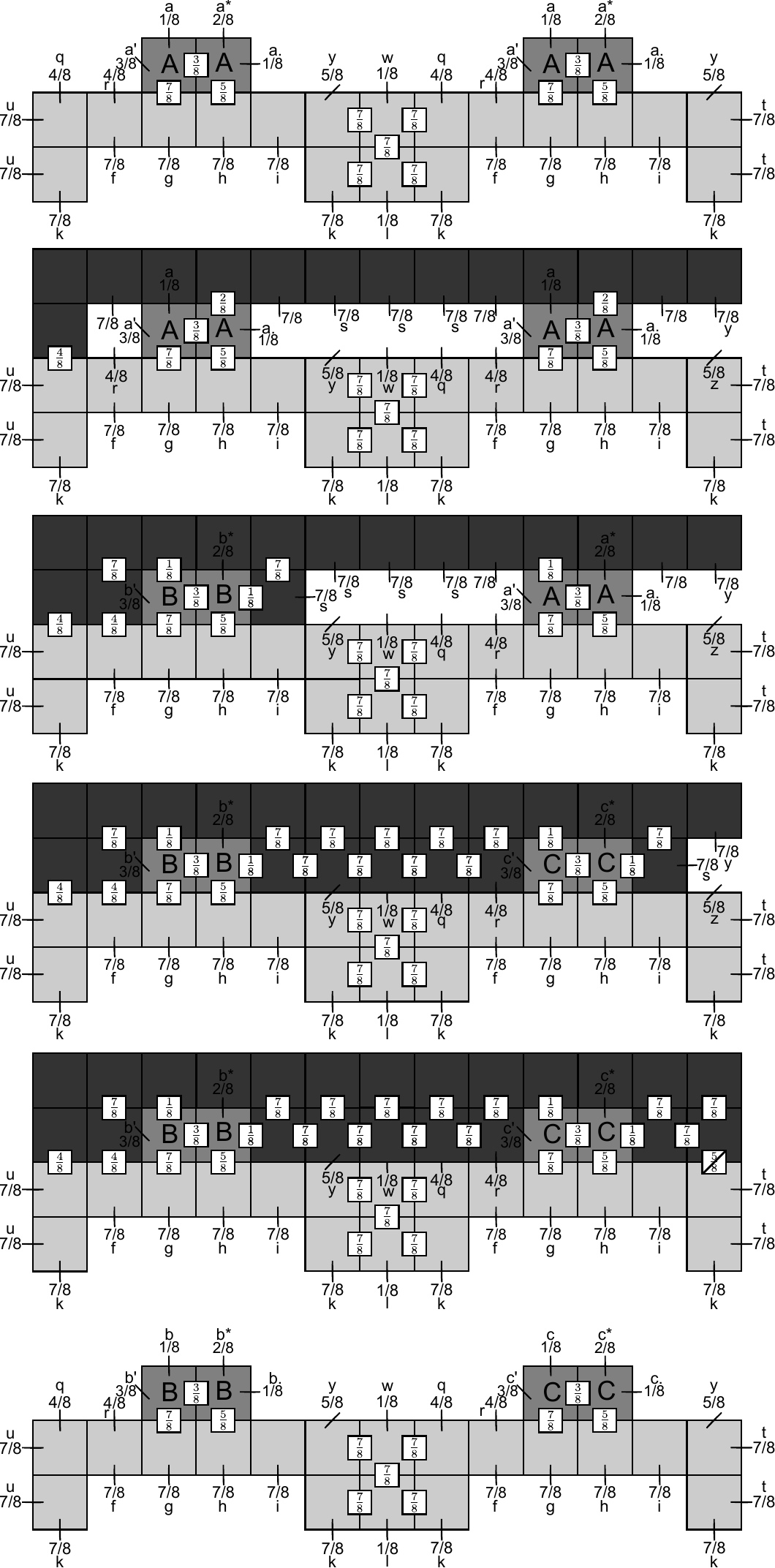}
	\caption{ This is an example of a transition gadget changing two As into an B and a C.
	\label{fig:transitionex} }
	\end{center}
\end{figure}

%\begin{figure}[htp]
%	\begin{center}
%	\includegraphics[scale=1.3]{Images/TuringMachine/transitionexsmall}
%	\caption{ This is the final state before the transition gadget and junk detaches.
%	\label{fig:transitionexsmall} }
%	\end{center}
%\end{figure}

Now in order to create the transition gadgets which represents a Turing instruction we must properly set the label of several unlabeled glues. We will temporarily number the unlabeled glues in Figure \ref{fig:turinginstruction} from $1$ starting at the left to $8$ at the farthest right. It does not matter whether you are moving the head left or right, the stateful pair always becomes stateless and the stateless pair always becomes stateful. Therefore, the only difference between moving the head right or left is that the stateful pair is at the left when moving right or at the right when moving left. We will continue assuming we are creating the transition gadgets for some $I$ which moves the head to the right. Since we are moving right, glues $3$ and $1$ will use the proper glue labels for the state symbol pair $(c, s)$. Glues $2$ and $4$ use the proper glue labels from the state symbol pairs $(\varnothing, p)$. Now, the next part explains the reason why we need $|\Gamma|$ gadgets. All gadgets now have glues $1$ through $4$ filled in with labels from our assembly set. Next fill in glues $5$ and $7$ for all gadgets with the proper glue labels from every state symbol pair that contains the $\varnothing$ state. Finally glues $6$ and $8$ are filled using the glue labels for every state symbol pair that contains state $n$ such that the symbol represented in $5$ and $7$ matches those of $6$ and $8$ within the same transition gadget.  Figure \ref{fig:transitionex} contains an overview of the necessary steps that allow the transition transition gadget to work.

With the gadget just about to detach, we can use the same arguments presented in Section \ref{sec:ocorrect} to prove that the transition gadget detaches correctly. The tape cell and transition gadget themselves are strongly connected and should not fall apart. Therefore, we start by saying that either both the tape cell and transition gadget end up on the same side of the cut or they end up on different sides of the cut. Using this information and the fact that we want a cut below $1$ we can whittle down the possible sides of all the tiles in either situation. As one will be able to see the only possible cut below $1$ happens when the tape cell and transition gadget are on opposite sides of the cut and all the fuel tiles are on the same side as the transition gadget.

%%%%%%%%%%%%%%%%%%%%%%%%%%%%%%%%%%%%%%%%%%%%%%%%%%%%%%%%%%%%%%%%%%%%%%%%%%%%%%%%%%%%%%%%%%%%%%%%%%%%%%%%%%%%%%%

\begin{figure}[htp]
	\begin{center}
	\includegraphics[scale=1.0]{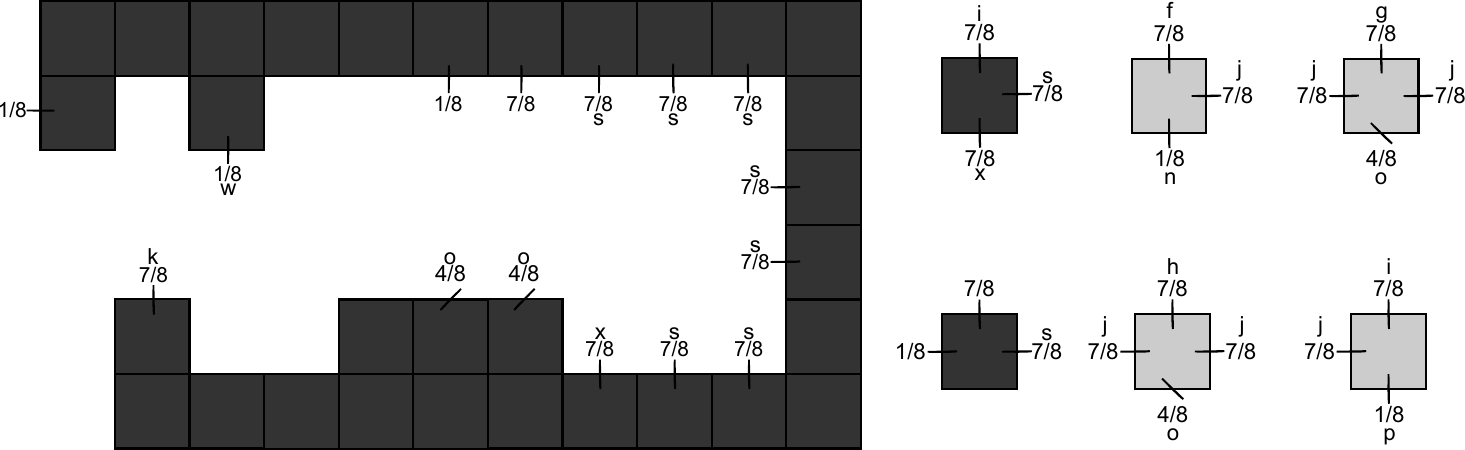}
	\caption{ The assembly needed to extend the tape to the right including two utility tiles colored dark gray. The four light gray tiles are necessary to allow the tape extension gadgets to remove themselves and not continuously attach. For extending to the left see Figure \ref{fig:extendleft} which is in the Appendix.
	\label{fig:extendright} \label{fig:tapeextdetach} }
	\end{center}
\end{figure}

\begin{figure}[htp]
	\begin{center}
	\includegraphics[scale=1.0]{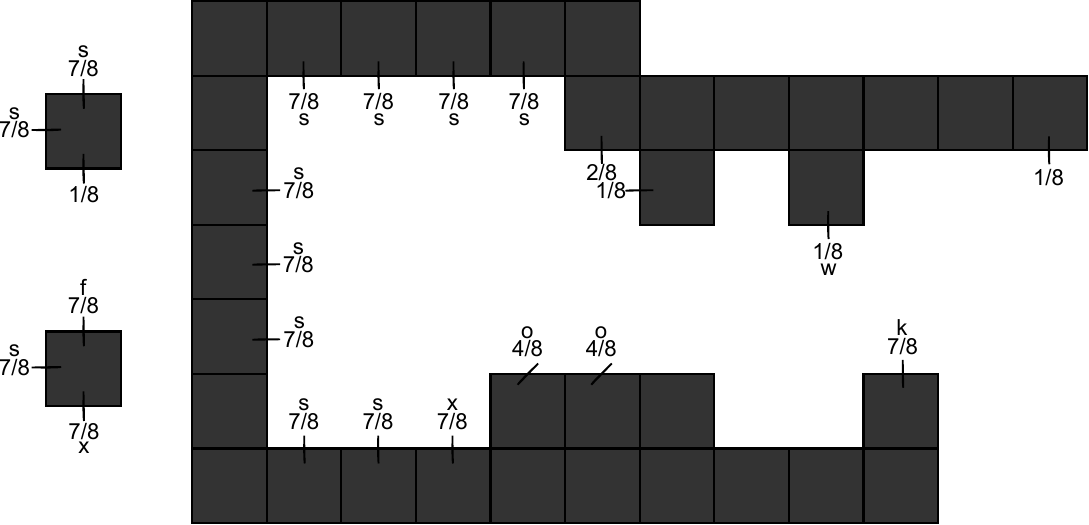}
	\caption{ The assembly needed to extend the tape to the left including two utility tiles.
	\label{fig:extendleft} }
	\end{center}
\end{figure}

\paragraph{Tape Extension.} \label{sec:tapeextension}  In order to simulate a Turing Machine and be space efficient we must ensure that our tape does not grow infinitely but extends only when needed. Our construction ensures that our tape grows only when necessary. We use a recurring theme that has been presented throughout the paper so far which is to place what we need and then use some negative glues to destabilize the then junk assembly. Due to some symmetric differences tape extension to the left is a bit different from that of the right but both work identically. The similarity between the two extension gadgets allows us to simply examine a single one and so only the right extension gadget will be discussed.

\begin{figure}[htp]
	\begin{center}
	\includegraphics[scale=.7]{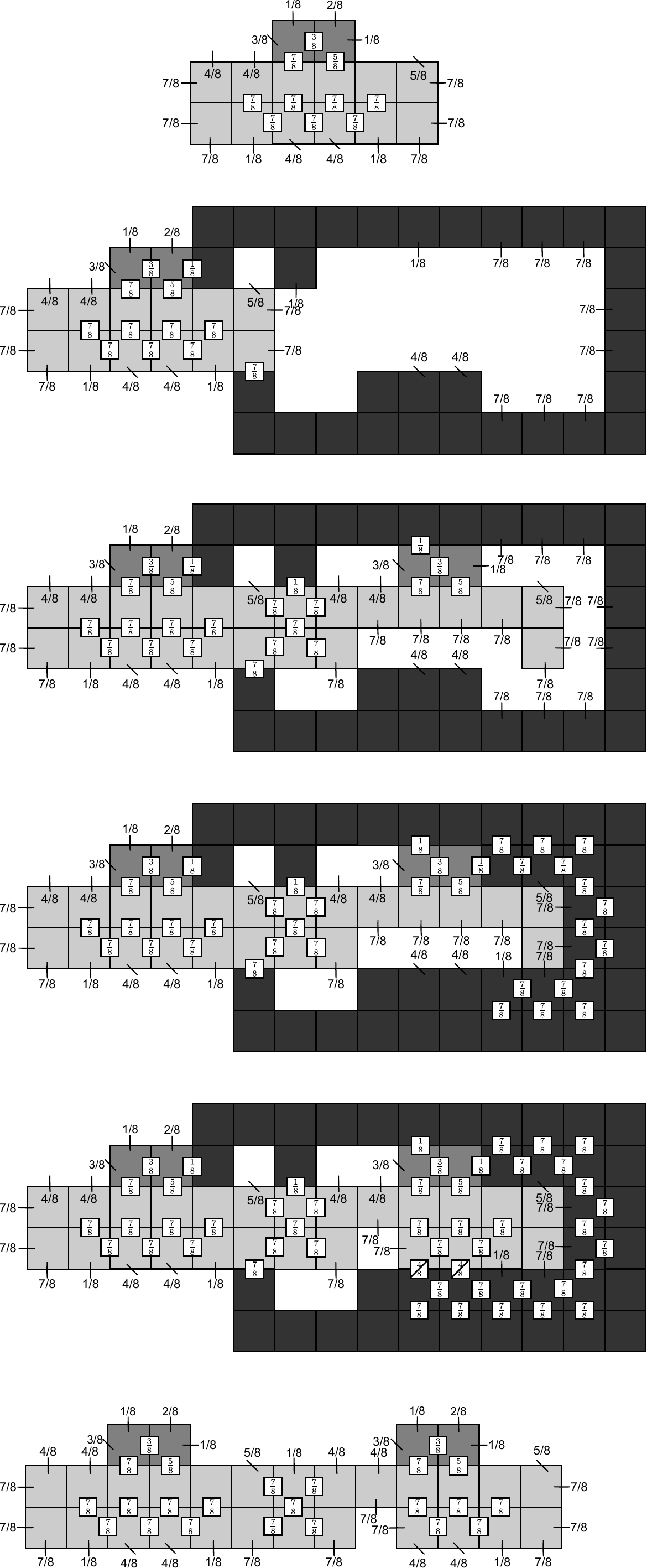}
	\caption{ This figure shows the steps that occur when extending the tape to the right.
	\label{fig:extendexp} }
	\end{center}
\end{figure}

We will need four tiles, shown in Figure \ref{fig:tapeextdetach}, which will serve as a way for the tape extension gadgets to detach themselves once they are done with their work. For simplicity we create a tape extension gadget for every stateful state symbol pair. Since it is only a single tile that needs to change in every tape extension gadget we could separate that single tile and just create a single tile for every stateful state symbol pair. The tape extension gadget can only attach if the position of the head is at the edge of the tape i.e. the stateful state symbol pair is at the edge. Once the head makes its way to the edge of the tape the tape extension gadget is able to attach and extend the tape. Then the tape extension gadget places the two tiles shown in Figure \ref{fig:tapecell} which will allow another tape cell assembly to attach. Once the new tape cell assembly has attached the stateless blank or ``default" state symbol pair can then be attached to the tape cell. After the attachment of the state symbol pair, tiles cooperatively bind around the tape extension arch until it attaches a tile underneath the tape cell arch. Finally, three tiles attach which cause the detachment of the extension gadget. A detailed overview of this process can be seen in Figure \ref{fig:extendexp}. The proof that the tape extension gadget detaches properly is easily verifiable at this point and is left as an exercise to the reader.

%\begin{figure}[htp]
%	\begin{center}
%	\includegraphics[scale=.6]{Images/TuringMachine/TapeExtension/extendexpsmall}
%	\caption{ This is the final step before the tape extension gadget and junk detaches. See Figure \ref{fig:extendexp} for a more detailed example.
%	\label{fig:extendexpsmall} }
%	\end{center}
%\end{figure}

%%%%%%%%%%%%%%%%%%%%%%%%%%%%%%%%%%%%%%%%%%%%%%%%%%%%%%%%%%%%%%%%%%%%%%%%%%%%%%%%%%%%%%%%%%%%%%%%%%%%%%%%%%%%%%%

\subsection{Tape Reduction} \label{tapereduction}

\begin{figure}[htp]
	\begin{center}
	\includegraphics[scale=1.0]{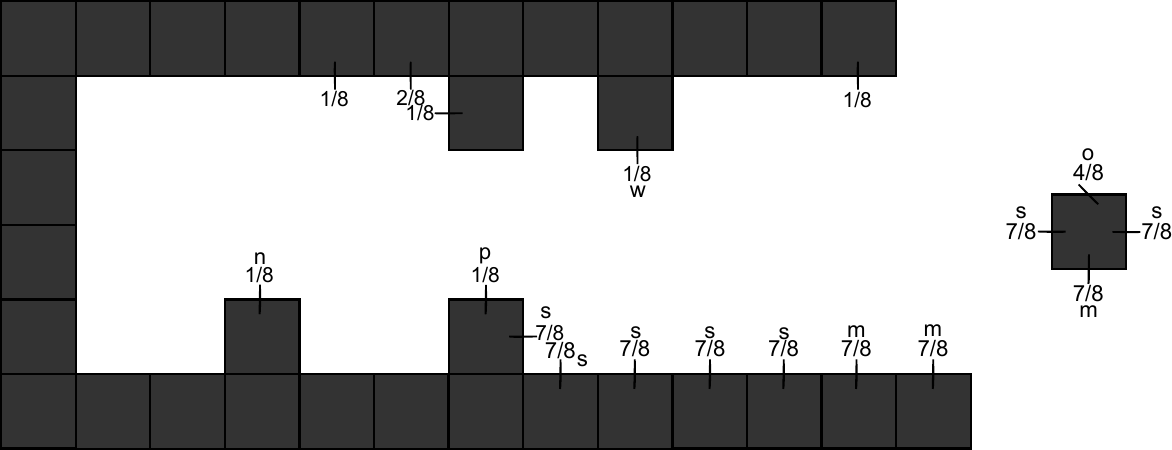}
	\caption{ The tape reduction gadget and extra tile needed to remove a single tape cell from the left edge.  This gadget relies on other tiles that have been presented in this paper.
	\label{fig:lshrinkgadget} }
	\end{center}
\end{figure}

%%%% Still need the right gadget

\begin{figure}[htp]
	\begin{center}
	\includegraphics[scale=1.0]{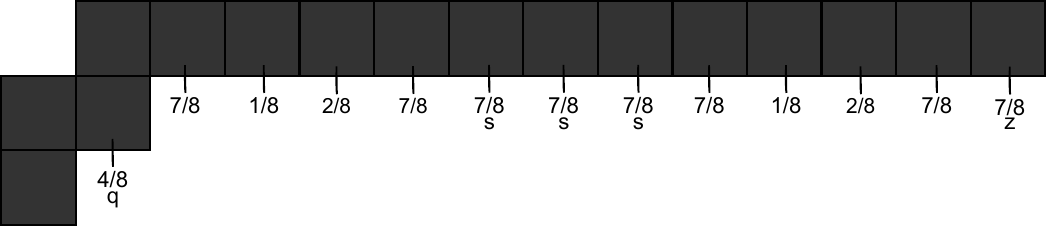}
	\caption{ This is an example of how a tail would be placed on a transition gadget that will ``blank" the left bit and move right.
	\label{fig:shrinktail} }
	\end{center}
\end{figure}

\begin{figure}[htp]
	\begin{center}
	\includegraphics[scale=.6]{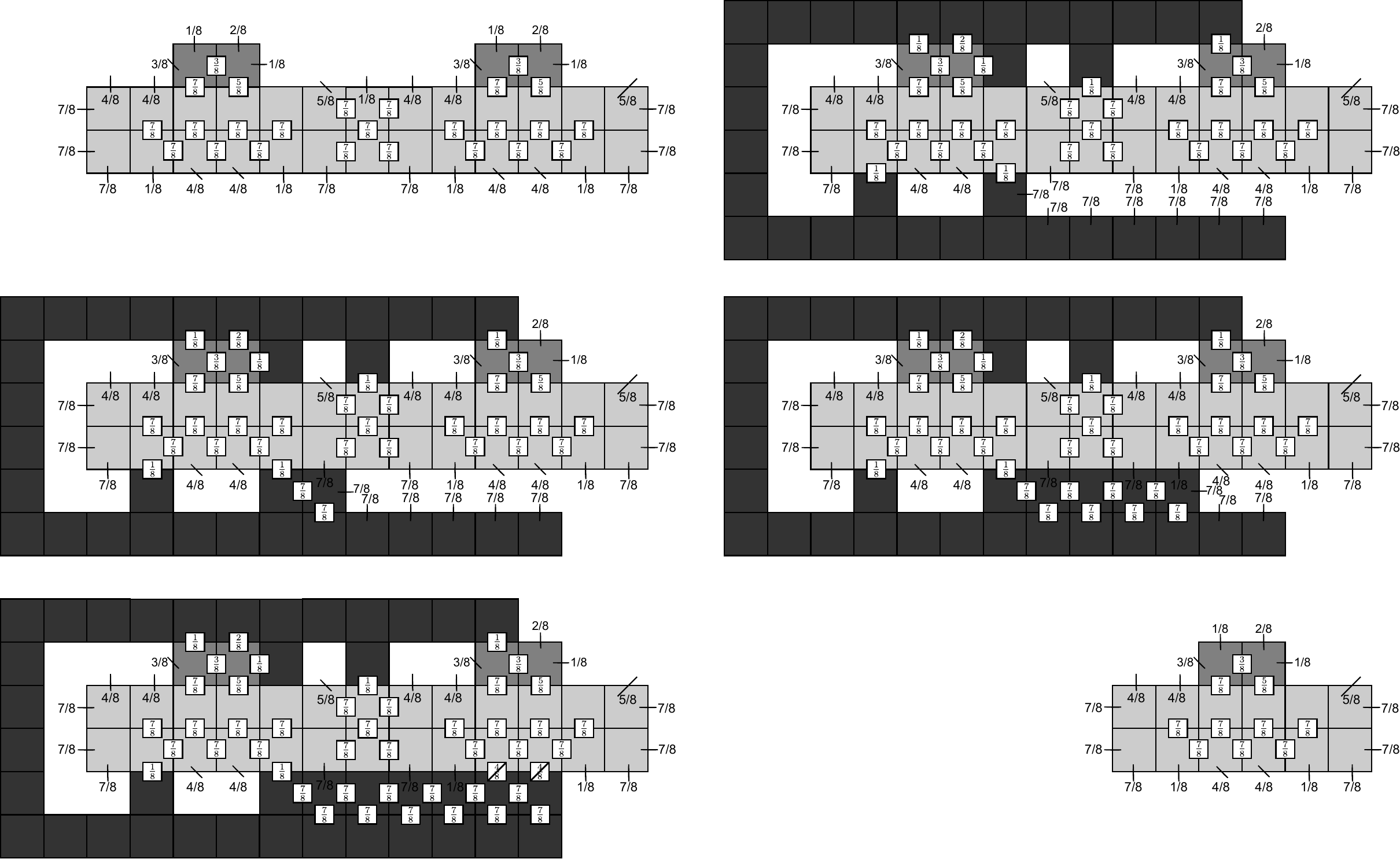}
	\caption{ This shows how a piece of tape will use the tape reduction gadget to shrink by one cell at the left end.
	\label{fig:shrinkex} }
	\end{center}
\end{figure}

A Turing machine in theory has an infinite length tape to work with.  To simulate this property our construction automatically extends the finite length tape on demand when needed.  However, it is reasonable that a Turing machine may grow a large tape, and then subsequently ``blank" out much of the tape leaving only a small length portion of tape between the leftmost and rightmost non-blank symbol.  In some sense, it is a waste of space to explicitly encode the large number of blank bits that are beyond this region. 

To not waste this space requires the inclusion of an additional mechanism into our construction, the \emph{tape reduction} gadget.  This gadget simply clips off the left or rightmost bit of the tape under certain circumstances such as if the leftmost or rightmost bit has been blanked.  This gadget allows for our Turing machine to achieve \emph{space efficiency}, which is the requirement that the size of the assembly representing each computational step is asymptotically bounded above by the size of the region of the tape between the leftmost and rightmost non-blank symbol, or the region between the head position and either the right or leftmost non-blank symbol.

One motivation for achieving space-efficiency is the potential implementation of components from active self-assembly models within our passive tile assembly model.  In particular, our space efficient construction may be utilized as the internal workings of a larger active component.  While the computation required of an active component may require a large tape size, the final output may be taken from a small set of options such as ``turn on glue a", or something similar.  With space efficiency, there is hope that our passive implementation will be able to perform the required functionality without growing so large that the geometry of the mechanism interferes with the higher-level active system mechanics.

Figure \ref{fig:lshrinkgadget} shows the gadget that is necessary to shrink the tape from the right.  One end of this gadget connects to a ``blank" bit while the other end connects to a bit and state combination that will then continue to blank the bit the head is at. An example of tape reduction can be seen in Figure \ref{fig:shrinkex} which shows an abridged set of steps necessary to make the tape shrink on the left end.  Without any way to synchronize the shrinking of the tape with the movement of the head one may still end up with many sequential ``blank" bits at the end of the tape.  In order to prevent this, whenever a transition gadget wants to ``blank" a bit on some end of the tape one must add a tail, such as the one shown in Figure \ref{fig:shrinktail}, to the transition gadget.  This tail must be added on the side that will be ``blanked" so as to be sure that all blank bits have been removed by tape reduction gadgets.

%In order to maintain space efficiency, which requires that the size of the assembly representing each computational step is asymptotically bounded above by the region of the tape between the leftmost and rightmost non-blank symbol, we have created tape reduction gadget(s). For clarity of presentation, and due to limited space, we do not present the full details of our extension gadget(s) in this extended abstract.

%%%%%%%%%%%%%%%%%%%%%%%%%%%%%%%%%%%%%%%%%%%%%%%%%%%%%%%%%%%%%%%%%%%%%%%%%%%%%%%%%%%%%%%%%%%%%%%%%%%%%%%%%%%%%%%

%\subsection{Turing Machine Implementation}
%Creating a specific Turing Machine requires that you create a transition gadget for every Turing Instruction in the Turing Machine, a state symbol set for every state and symbol combination, tape cell assemblies, and finally all the assemblies necessary for extending the tape. Should you wish to have a non-deterministic Turing Machine you would simply add multiple transition gadgets for whichever bit state combinations you wanted to have a non-deterministic instruction. One important thing to keep in mind is that if you have transition gadget which takes as arguments two stateless state symbol pairs it will produce undefined behavior.

%\input{conclusion}

%\clearpage
%\appendix
%\noindent {\bf\huge Technical appendix}
%\input{appendix-dummySec}

\bibliographystyle{amsplain}
\bibliography{tam,tamUSA}

\end{document}